# Improved Sparsity Thresholds Through Dictionary Splitting

Patrick Kuppinger, Giuseppe Durisi, and Helmut Bölcskei
ETH Zurich, 8092 Zurich, Switzerland
E-mail: {patricku, gdurisi, boelcskei}@nari.ee.ethz.ch

*Abstract*—Known sparsity thresholds for basis pursuit to deliver the maximally sparse solution of the compressed sensing recovery problem typically depend on the dictionary's coherence. While the coherence is easy to compute, it can lead to rather pessimistic thresholds as it captures only limited information about the dictionary. In this paper, we show that viewing the dictionary as the concatenation of two general sub-dictionaries leads to provably better sparsity thresholds—that are explicit in the coherence parameters of the dictionary and of the individual sub-dictionaries. Equivalently, our results can be interpreted as sparsity thresholds for dictionaries that are unions of two general (i.e., not necessarily orthonormal) sub-dictionaries.

## I. INTRODUCTION

A typical problem in compressed sensing (CS) [1], [2] is to find the sparsest representation of a given vector as a linear combination of vectors from a given set, commonly referred to as *dictionary* [3]–[9]. More concretely, let a dictionary be a set of $N$ elements $\mathbf{d}_i \in \mathbb{C}^M$ ($i = 1, \ldots, N$), which we assume (throughout the paper) to have unit $\ell_2$-norm and to span $\mathbb{C}^M$. We organize the dictionary elements into an $M \times N$ matrix $\mathbf{D} = [\mathbf{d}_1 \ldots \mathbf{d}_N]$; typically $M \ll N$. Given the vector $\mathbf{y} \in \mathbb{C}^M$, the problem of finding its sparsest representation in $\mathbf{D}$ can be stated as follows:

(P0)  find $\arg\min \|\mathbf{x}\|_0$  subject to $\mathbf{y} = \mathbf{D}\mathbf{x}$.

Here, $\|\mathbf{x}\|_0$ denotes the number of nonzero entries of the vector $\mathbf{x}$. Due to the combinatorial nature of the problem, solving (P0) is infeasible for practically relevant problem sizes $N, M$. A standard approach in the CS literature is to relax (P0) to the following optimization problem, commonly referred to as *basis pursuit* [4]–[9]

(P1)  find $\arg\min \|\mathbf{x}\|_1$  subject to $\mathbf{y} = \mathbf{D}\mathbf{x}$

where $\|\mathbf{x}\|_1$ denotes the $\ell_1$-norm of $\mathbf{x}$, and to ask under which conditions the solutions of (P1) and (P0) coincide. The advantage of this approach resides in the fact that (P1) can be cast into a linear program [4] and can therefore be solved (more) efficiently.

It turns out that the solutions of (P0) and (P1) are unique if there exists a vector $\mathbf{x}$ that satisfies $\mathbf{y} = \mathbf{D}\mathbf{x}$ with $\|\mathbf{x}\|_0$ less than a certain *sparsity threshold* [5]–[9] (in the following often referred to simply as threshold). This sparsity threshold is a function of the dictionary's *coherence* [5], defined as the maximum of the absolute value of the inner product between pairs of distinct columns of $\mathbf{D}$, where the maximum is taken over all such pairs in the dictionary. It was furthermore shown in [8] that a sparsity threshold that guarantees the uniqueness of the solution of (P1) necessarily guarantees that this solution is also the unique solution of (P0).

The coherence is easy to compute but often leads to rather pessimistic thresholds as it captures only limited information about the dictionary. Using additional structural information about the dictionary can lead to improved (i.e., higher) thresholds. For example, in [7], [10] sparsity thresholds for a dictionary that consists of the union of two orthonormal bases (ONBs) are derived that are about a factor of 2 higher than the corresponding thresholds for a general dictionary with the same coherence.

*Contributions:* We consider dictionaries—with coherence $d$—that consist of two general sub-dictionaries with coherence $a$ and $b$, respectively. The two sub-dictionaries need not be ONBs, need not have the same number of elements, and do not need to span $\mathbb{C}^M$. We denote the set of all such dictionaries as $\mathcal{D}(d, a, b)$;[1] the individual dictionaries in $\mathcal{D}(d, a, b)$ need not have the same cardinality.

For dictionaries in $\mathcal{D}(d, a, b)$, we provide a sparsity threshold that guarantees uniqueness of the solution of (P0), and a threshold that guarantees uniqueness of the solution of (P1) and, hence, equivalence of the solutions of (P1) and (P0). These thresholds depend on the parameters $d$, $a$, and $b$ only and improve on the threshold that would be obtained if one treated the union of two sub-dictionaries as a general dictionary with coherence $d$ and applied the threshold in [6], [8]. The improvement is substantial whenever $a, b \ll d$. The thresholds found in [7], [10] for the two-ONB setting are special cases of our thresholds. We furthermore show that thinking of a general dictionary as the concatenation of two (general) sub-dictionaries, the thresholds derived in this paper strictly improve upon the one reported in [6], [8], unless $b = d$ in which case our thresholds coincide with the one in [6], [8]. Any such *splitting of a dictionary* will yield improved sparsity thresholds. However, finding the optimal split, for a given dictionary, in terms of obtaining the best possible thresholds is a combinatorial problem.

Our threshold for uniqueness of the solution of (P0) is based on a generalization of the uncertainty relation for the two-ONB setting [7] to dictionaries in $\mathcal{D}(d, a, b)$. The derivation of the threshold for the uniqueness of the solution of (P1) is a generalization of the corresponding derivation for the two-ONB case reported in [10].

*Notation:* Throughout the paper, we use lowercase boldface letters for column vectors, e.g., $\mathbf{x}$, and uppercase boldface letters for matrices, e.g., $\mathbf{D}$. For a given matrix $\mathbf{D}$, we denote its conjugate transpose by $\mathbf{D}^H$, and the element in its $i$th row and $j$th column by $[\mathbf{D}]_{i,j}$; furthermore, $\mathbf{d}_i$ stands for its $i$th column, and $\mathrm{kern}(\mathbf{D})$ is its null space. The $i$th element of a vector $\mathbf{x}$ is denoted by $x_i$. The Euclidean or $\ell_2$-norm of the vector $\mathbf{x}$ is $\|\mathbf{x}\|_2 = \sqrt{\mathbf{x}^H \mathbf{x}}$, the $\ell_1$-norm is defined as $\|\mathbf{x}\|_1 = \sum_i |x_i|$, and $\|\mathbf{x}\|_\infty = \max_i |x_i|$ is the $\ell_\infty$-norm of $\mathbf{x}$. The smallest eigenvalue of the positive-definite matrix $\mathbf{G}$ is denoted by $\lambda_{\min}(\mathbf{G})$. For $u \in \mathbb{R}$, we define $[u]^+ = \max\{0, u\}$.

---
[1]Without loss of generality, we assume $a \leq b$ in the remainder of the paper.

## II. Previous Results

In this section, we review relevant sparsity thresholds [5]–[10].

The *spark* of a dictionary $\mathbf{D}$, defined as the cardinality of the smallest set of linearly dependent columns of $\mathbf{D}$ [6], sheds light on the conditions for uniqueness of the solution of (P0). In particular, the following result holds: for a given dictionary $\mathbf{D}$, a vector[2] $\mathbf{x} \in \mathbb{C}^N$ is the unique solution of (P0) if and only if [6], [8]

$$\|\mathbf{x}\|_0 < \mathrm{spark}(\mathbf{D})/2$$

where $\mathrm{spark}(\mathbf{D})$ denotes the spark of $\mathbf{D}$. Determining the spark of a dictionary is a combinatorial problem and therefore computationally expensive, in general. A common approach to overcome this problem is to derive lower bounds on $\mathrm{spark}(\mathbf{D})$ that are explicit in an easy-to-compute parameter of the dictionary. Let, for example, the coherence of a dictionary $\mathbf{D}$ be defined as[3]

$$d = \max_{i \neq j} |\mathbf{d}_i^H \mathbf{d}_j| \quad (1)$$

and let $\mathcal{D}_{\mathrm{gen}}(d)$ be the set of all dictionaries with coherence $d$. It is shown in [6], [8], [9] that for any dictionary $\mathbf{D}$ in $\mathcal{D}_{\mathrm{gen}}(d)$ we have that[4]

$$\mathrm{spark}(\mathbf{D}) \geq 1 + 1/d. \quad (2)$$

Consequently, for a dictionary with coherence $d$, a sufficient condition for a vector $\mathbf{x}$ to be the unique solution of (P0) is

$$\|\mathbf{x}\|_0 < \frac{1}{2}\left(1 + \frac{1}{d}\right). \quad (3)$$

It is furthermore shown in [6], [8] that when (3) holds, the solution of (P1) is unique as well (and hence equal to the unique solution of (P0)).

The threshold on the right-hand side (RHS) of (3) applies to any dictionary in $\mathcal{D}_{\mathrm{gen}}(d)$. This set is, however, rather large and the dictionaries within the set can differ widely in their structure: from equiangular tight frames [11], where each pair of elements in the dictionary achieves the maximum in (1), to dictionaries where the maximum is achieved by one pair only. For a given dictionary, the threshold on the RHS of (3) might therefore be rather pessimistic.

Better sparsity thresholds can be obtained if structural properties of the dictionary are explicitly taken into account. This leads to sparsity thresholds that apply to smaller sets of dictionaries. Let, for example, $\mathcal{D}_{\mathrm{onb}}(d) \subset \mathcal{D}_{\mathrm{gen}}(d)$ be the set of dictionaries with coherence $d$ that consist of the union of two ONBs. For any dictionary in $\mathcal{D}_{\mathrm{onb}}(d)$, a sufficient condition for $\mathbf{x}$ to be the unique solution of (P0) is [7], [8], [10]

$$\|\mathbf{x}\|_0 < 1/d. \quad (4)$$

[2]In the remainder of the paper, whenever we speak of a vector $\mathbf{x}$, we mean that this vector is consistent with the observation $\mathbf{y}$, i.e., it satisfies $\mathbf{y} = \mathbf{D}\mathbf{x}$.
[3]For dictionaries consisting of only one element we set $d = 0$.
[4]Note that $d \leq 1$ as we assumed that $\|\mathbf{d}_i\|_2 = 1$, $i = 1, \ldots, N$.

A sufficient condition for $\mathbf{x}$ to be the unique solution of (P1) (which hence equals the unique solution of (P0)) is [7], [8], [10]

$$\|\mathbf{x}\|_0 < (\sqrt{2} - 0.5)/d. \quad (5)$$

For small $d$, the thresholds in (4) and (5) are almost a factor of 2 larger than the threshold on the RHS of (3). Surprisingly, the threshold in (5) drops below that in (3) for $d > 2(\sqrt{2} - 1)$.

## III. Two Novel Sparsity Thresholds

We focus on the class $\mathcal{D}(d, a, b)$ of dictionaries with coherence $d$ that consist of two sub-dictionaries with coherence $a$ and $b$, respectively. A generic dictionary in $\mathcal{D}(d, a, b)$ is denoted as $\mathbf{D} = [\mathbf{A}\ \mathbf{B}]$, where $\mathbf{A}$ has coherence $a$ and $\mathbf{B}$ has coherence $b$. The sub-dictionaries $\mathbf{A}$ and $\mathbf{B}$ need not be ONBs, need not have the same number of elements, and do not need to span $\mathbb{C}^M$. Note that $\mathcal{D}(d, a, b) \subset \mathcal{D}_{\mathrm{gen}}(d)$; hence, the sparsity threshold in (3) applies to any dictionary in $\mathcal{D}(d, a, b)$ as well. Furthermore, we also have that[5] $\mathcal{D}_{\mathrm{onb}}(d) \subset \mathcal{D}(d, 0, 0)$. In Theorem 1 and 2 below, we derive sparsity thresholds that apply to the set $\mathcal{D}(d, a, b)$, are functions of $d$, $a$, and $b$ only, and improve on the threshold that would be obtained if one treated the union of two sub-dictionaries as a general dictionary with coherence $d$. The basis for our result on the uniqueness of the solution of (P0) is an uncertainty relation for dictionaries in $\mathcal{D}(d, a, b)$ stated next.

*Lemma 1:* Let $\mathbf{D} = [\mathbf{A}\ \mathbf{B}]$, $\mathbf{D} \in \mathbb{C}^{M \times N}$, be a dictionary in $\mathcal{D}(d, a, b)$. For any vector in $\mathbb{C}^M$ that can be represented as a linear combination of $n_a$ linearly independent columns of $\mathbf{A}$ and, equivalently, as a linear combination of $n_b$ linearly independent columns of $\mathbf{B}$, the following inequality holds:

$$n_a n_b \geq [1 - a(n_a - 1)]^+ [1 - b(n_b - 1)]^+ / d^2. \quad (6)$$

*Proof:* See Appendix A. ∎

Note that for $a = b = 0$, (6) reduces to $n_a n_b \geq 1/d^2$, which is the well-known uncertainty relation for the union of two ONBs reported in [7, Th. 1].

We next provide a lower bound on the spark of dictionaries in the set $\mathcal{D}(d, a, b)$.

*Theorem 1:* The spark of any dictionary $\mathbf{D}$ in the set $\mathcal{D}(d, a, b)$ is lower-bounded according to

$$\mathrm{spark}(\mathbf{D}) \geq \hat{x} + f(\hat{x})$$

where

$$f(x) = \frac{(1 + a)(1 + b) - xb(1 + a)}{x(d^2 - ab) + a(1 + b)}$$

and

$$\hat{x} = \min\{x_b, x_s\}$$

with

$$x_b = (1 + b)/(b + d^2)$$

[5]$\mathcal{D}(d, 0, 0)$, for example, also contains dictionaries that consist of an ONB and an orthonormal set which does not form a basis.

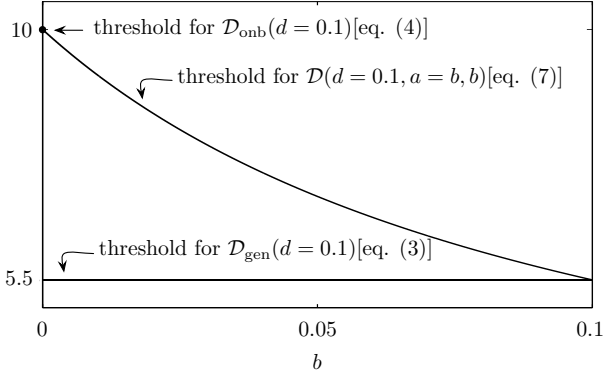

Figure 1. Sparsity thresholds for dictionaries in $\mathcal{D}_{\text{gen}}(d)$, $\mathcal{D}_{\text{onb}}(d)$, and $\mathcal{D}(d,a,b)$. The case $d=0.1$ and $a=b$ is considered. Note that when $a=b$, the bound (7) reduces to $\|\mathbf{x}\|_0 < (1+b)/(d+b)$.

and

$$x_s = \begin{cases} 1/d, & \text{if } a = b = d, \\ \dfrac{d\sqrt{(1+a)(1+b)} - a - ab}{d^2 - ab}, & \text{otherwise.} \end{cases}$$

*Proof:* See Appendix B. ∎

Theorem 1 yields the following sufficient condition for a vector $\mathbf{x}$ to be the unique solution of (P0) for any $\mathbf{D} \in \mathcal{D}(d,a,b)$:

$$\|\mathbf{x}\|_0 < \bigl[\hat{x} + f(\hat{x})\bigr]/2. \tag{7}$$

The threshold in (7) reduces to the one in (3) when $a = b = d$, or when $x_b \leq x_s$ and $b = d$. In all other cases, the threshold in (7) is strictly larger than that in (3), as shown in Appendix C. Furthermore, (7) reduces to (4) when $a = b = 0$. As illustrated in Fig. 1 for the case $a = b$, the improvement of the threshold in (7) over that in (3) is significant when $a, b \ll d$.

We next present a sparsity threshold that guarantees the uniqueness of the solution of (P1) (which is hence equal to the unique solution of (P0)) for any dictionary in $\mathcal{D}(d,a,b)$.

*Theorem 2:* For any dictionary $\mathbf{D}$ in $\mathcal{D}(d,a,b)$ a sufficient condition for (P1) to have a unique solution is

$$\|\mathbf{x}\|_0 < \begin{cases} \dfrac{\delta\bigl[\epsilon - (d+3b)\bigr]}{2(d^2 - b^2)}, & \text{if } b < d \text{ and } \kappa(d,b) > 1, \\[6pt] \dfrac{1 + 2d^2 + 3b - d\delta}{2(d^2 + b)}, & \text{otherwise} \end{cases} \tag{8}$$

with

$$\kappa(d,b) = \frac{\delta\sqrt{2d\,(b+3d+\epsilon)} - 2d - 2b(\delta + d)}{2(d^2 - b^2)}$$

where $\delta = 1 + b$ and $\epsilon = 2\sqrt{2}\sqrt{d(b+d)}$.

*Proof:* See Appendix D. ∎

The threshold in (8) depends on $b$ and $d$ only. An improved threshold, which depends on $a$ as well, can be obtained through minor modifications of the proof in Appendix D. The resulting expression is, however, unwieldy.

The threshold in (8) reduces to the one in (3) whenever $b = d$ (independently of the value of $a$). Otherwise, the threshold in (8) is strictly larger than that in (3). Hence, viewing a given dictionary $\mathbf{D}$ as the concatenation of two sub-dictionaries $\mathbf{A}$ and $\mathbf{B}$ (i.e., splitting the dictionary $\mathbf{D}$ into $\mathbf{A}$ and $\mathbf{B}$) yields improved sparsity thresholds, as a function of $d$, $a$, and $b$, compared to the threshold $(1 + 1/d)/2$ in [6], [8].

For $a = b = 0$, the threshold in (8) reduces to

$$\|\mathbf{x}\|_0 < \begin{cases} (\sqrt{2} - 0.5)/d, & \text{if } d < 1/\sqrt{2}, \\ 1 + (1-d)/(2d^2), & \text{otherwise.} \end{cases} \tag{9}$$

Note that (9) coincides with the two-ONB threshold in (5) when $d < 1/\sqrt{2}$, but improves on (5) for $d \geq 1/\sqrt{2}$. Furthermore, differently from the threshold in (5), the threshold in (9) cannot fall below the threshold for $\mathcal{D}_{\text{gen}}(d)$ given by $(1 + 1/d)/2$.

## APPENDIX A
## PROOF OF LEMMA 1

Let $\mathbf{y}$ be a vector in $\mathbb{C}^M$ that can be represented as a linear combination of $n_a$ linearly independent columns of $\mathbf{A}$ and, equivalently, as a linear combination of $n_b$ linearly independent columns of $\mathbf{B}$. Let $\mathcal{P}$ and $\mathcal{Q}$ be the sets containing the indices of the participating columns in $\mathbf{A}$ and $\mathbf{B}$, respectively. Then $\mathbf{y} = \sum_{i \in \mathcal{P}} p_i \mathbf{a}_i = \sum_{j \in \mathcal{Q}} q_j \mathbf{b}_j$ for some scalars $\{p_i\}$ and $\{q_j\}$. Let $\mathbf{p}$ be the $n_a$-dimensional vector containing the scalars $\{p_i\}$, and, similarly, let $\mathbf{q}$ be the $n_b$-dimensional vector containing the scalars $\{q_j\}$. The following chain of inequalities holds:

$$\|\mathbf{y}\|_2^2 = \sum_{i \in \mathcal{P}} \sum_{j \in \mathcal{Q}} p_i^* \mathbf{a}_i^H \mathbf{b}_j q_j \leq \sum_{i \in \mathcal{P}} \sum_{j \in \mathcal{Q}} |p_i|\,|q_j|\,\bigl|\mathbf{a}_i^H \mathbf{b}_j\bigr|$$
$$\leq d\|\mathbf{p}\|_1 \|\mathbf{q}\|_1 \leq d\sqrt{n_a}\|\mathbf{p}\|_2 \sqrt{n_b}\|\mathbf{q}\|_2. \tag{10}$$

Denote by $\mathbf{A}_\mathcal{P}$ the matrix containing the columns of $\mathbf{A}$ corresponding to the index set $\mathcal{P}$. Furthermore, let $\mathbf{G}_\mathcal{P} = \mathbf{A}_\mathcal{P}^H \mathbf{A}_\mathcal{P}$. Then, we can lower-bound the $\ell_2$-norm of $\mathbf{y}$ as follows:

$$\|\mathbf{y}\|_2^2 = \mathbf{p}^H \underbrace{\mathbf{A}_\mathcal{P}^H \mathbf{A}_\mathcal{P}}_{=\,\mathbf{G}_\mathcal{P}} \mathbf{p} \geq \lambda_{\min}(\mathbf{G}_\mathcal{P})\|\mathbf{p}\|_2^2. \tag{11}$$

Using analogous steps (and notation), we also get

$$\|\mathbf{y}\|_2^2 = \mathbf{q}^H \underbrace{\mathbf{B}_\mathcal{Q}^H \mathbf{B}_\mathcal{Q}}_{=\,\mathbf{G}_\mathcal{Q}} \mathbf{q} \geq \lambda_{\min}(\mathbf{G}_\mathcal{Q})\|\mathbf{q}\|_2^2. \tag{12}$$

Note that both $\lambda_{\min}(\mathbf{G}_\mathcal{P})$ and $\lambda_{\min}(\mathbf{G}_\mathcal{Q})$ are strictly positive, because the columns of $\mathbf{A}_\mathcal{P}$ and of $\mathbf{B}_\mathcal{Q}$ are linearly independent, by assumption. Inserting (11) and (12) into (10) yields

$$\sqrt{n_a n_b} \geq \frac{1}{d}\sqrt{\lambda_{\min}(\mathbf{G}_\mathcal{P})\lambda_{\min}(\mathbf{G}_\mathcal{Q})}. \tag{13}$$

Next, proceeding as in [9, Lem. 2.3], we bound $\lambda_{\min}(\mathbf{G}_\mathcal{P})$ and $\lambda_{\min}(\mathbf{G}_\mathcal{Q})$ from below using Geršgorin's disc theorem [12, Th. 6.1.1] and obtain $\lambda_{\min}(\mathbf{G}_\mathcal{P}) \geq \max\{0, 1 - a(n_a - 1)\}$ and $\lambda_{\min}(\mathbf{G}_\mathcal{Q}) \geq \max\{0, 1 - b(n_b - 1)\}$, respectively. The proof is concluded by inserting these two lower bounds into (13).

## APPENDIX B
## PROOF OF THEOREM 1

Let $\mathbf{D}$ be an arbitrary dictionary in $\mathcal{D}(d,a,b)$, and let $Z(\mathbf{D})$ be the cardinality of the smallest set of linearly dependent columns of $\mathbf{D}$ consisting of $n_a \geq 1$ linearly independent columns of $\mathbf{A}$ and $n_b \geq 1$ linearly independent columns of $\mathbf{B}$. By the definition of the spark [6, Def. 1], we have that

$$\text{spark}(\mathbf{D}) = \min\{\text{spark}(\mathbf{A}), \text{spark}(\mathbf{B}), Z(\mathbf{D})\}. \quad (14)$$

We next lower-bound each term in (14). From (2) we get

$$\text{spark}(\mathbf{A}) \geq 1 + 1/a \quad \text{and} \quad \text{spark}(\mathbf{B}) \geq 1 + 1/b.$$

Since $a \leq b$, by assumption, we obtain

$$\text{spark}(\mathbf{D}) \geq \min\{1 + 1/b, Z(\mathbf{D})\}. \quad (15)$$

We now seek a lower bound on the RHS of (15). To this end, we start by noting that $Z(\mathbf{D}) = n_a + n_b$. In the light of (15), we need to focus only on dictionaries that have $n_a + n_b \leq 1 + 1/b$. From the uncertainty relation in Lemma 1, we know that the product $n_a n_b$ cannot be arbitrarily small. More specifically, we have that $n_a n_b \geq [1 - a(n_a - 1)][1 - b(n_b - 1)]/d^2$. Solving for $n_a$, we get[6]

$$n_a \geq \frac{(1+a)(1+b) - n_b b(1+a)}{n_b(d^2 - ab) + a(1+b)} = f(n_b).$$

Finally, adding $n_b$ on both sides yields

$$Z(\mathbf{D}) = n_a + n_b \geq f(n_b) + n_b. \quad (16)$$

To obtain an expression that is explicit in $d$, $a$, and $b$ only, we further lower-bound the RHS of (16) by minimizing $f(n_b) + n_b$ over $n_b$, under the additional constraints $n_a \geq 1$ and $n_b \geq 1$ (implied by assumption). The resulting bound reads

$$Z(\mathbf{D}) \geq \min_{n_b \geq 1}[n_b + \max\{f(n_b), 1\}].$$

Since

$$\min_{n_b \geq 1}[n_b + \max\{f(n_b), 1\}] \leq [n_b + \max\{f(n_b), 1\}]\big|_{n_b = 1/b}$$
$$= 1 + 1/b$$

we can lower-bound the RHS of (14) by

$$\text{spark}(\mathbf{D}) \geq \min_{n_b \geq 1}[n_b + \max\{f(n_b), 1\}]$$
$$\geq \min_{x \geq 1}[x + \max\{f(x), 1\}]. \quad (17)$$

Here, in the last step, we replaced the integer parameter $n_b$ by the real-valued parameter $x$. We next compute the minimum in (17). The function $f(x)$ is monotonically decreasing. Furthermore, the equation $f(x) = 1$ has the unique solution $x_b = (1+b)/(b+d^2) \geq 1$, where the inequality follows because $d \leq 1$ by definition. We can, therefore, rewrite (17) as

$$\min_{x \geq 1}[x + \max\{f(x), 1\}] = \min_{1 \leq x \leq x_b}[x + f(x)]. \quad (18)$$

In the case $a = b = d$, the function $g(x) = x + f(x)$ reduces to the constant $1 + 1/d$ so that $\text{spark}(\mathbf{D}) \geq 1 + 1/d$. In all other cases, the function $g(x)$ is strictly convex for $x \geq 0$. Furthermore, we have $g(1) \geq g(x_b)$ as a consequence of the assumption $a \leq b$. Hence, the minimum in (18) is achieved—as a result of the convexity of $g(x)$—either at the boundary point $x_b$, or at the

---

[6]Equivalently, we could solve for $n_b$ and obtain the same final result.

stationary point $x_s$ of $g(x)$, which is given by

$$x_s = \frac{d\sqrt{(1+a)(1+b)} - a - ab}{d^2 - ab} \geq 1. \quad (19)$$

The inequality in (19) follows from the convexity of $g(x)$ and the fact that $g(1) \geq g(x_b)$.

## APPENDIX C

*The case $a = b = d$:* The threshold in (7) reduces to the well-known [6], [8] threshold $(1 + 1/d)/2$ in (3).

*The case $x_b \leq x_s$:* The threshold in (7) equals

$$\frac{\hat{x} + f(\hat{x})}{2} = \frac{x_b + f(x_b)}{2} = \frac{1}{2}\left(1 + \frac{1+b}{b+d^2}\right).$$

It is now easily verified that

$$\frac{1}{2}\left(1 + \frac{1+b}{b+d^2}\right) \geq \frac{1}{2}\left(1 + \frac{1}{d}\right)$$

for all $b \leq d \leq 1$ with equality only if $b = d$.

*The case $x_b > x_s$:* Set $\Delta = \sqrt{(1+a)(1+b)} - d$. The function $x_s + f(x_s)$, which depends on the variables $d$, $a$, and $b$, is strictly monotonically decreasing in $a$ as long as $b/d < \Delta < d/b$. The inequality $\Delta < d/b$ is always satisfied, because $a \leq b \leq d$. The inequality $b/d < \Delta$ holds whenever $x_s < x_b$, which is satisfied by assumption. Hence, we have that

$$\frac{x_s + f(x_s)}{2} \geq \frac{x_s + f(x_s)}{2}\bigg|_{a=b} = \frac{1+b}{d+b} \geq \frac{1}{2}\left(1 + \frac{1}{d}\right).$$

Here, equality holds only for the case $a = b = d$, already treated separately above.

## APPENDIX D
### PROOF OF THEOREM 2

The proof follows closely that of [10, Prop. 3], which deals with the case where $\mathbf{A}$ and $\mathbf{B}$ are ONBs. Our proof makes use of the following result.

*Proposition 1 ([8, Lem. 1, Lem. 2], [10, Prop. 2]):*
Let $S_L(\mathbf{v})$ denote the sum of the largest (in absolute value) $L$ entries of the vector $\mathbf{v} \in \mathbb{C}^N$ and assume that

$$P_L(\mathbf{D}) = \max_{\mathbf{v} \neq 0, \, \mathbf{v} \in \text{kern}(\mathbf{D})} \frac{S_L(\mathbf{v})}{\|\mathbf{v}\|_1} < \frac{1}{2}.$$

Then, for all $\mathbf{x} \in \mathbb{C}^N$ with $\|\mathbf{x}\|_0 < L$ and for all nonzero $\mathbf{v} \in \text{kern}(\mathbf{D})$ we have that $\|\mathbf{x} + \mathbf{v}\|_0 > \|\mathbf{x}\|_0$ and $\|\mathbf{x} + \mathbf{v}\|_1 > \|\mathbf{x}\|_1$.

Proposition 1 implies that $\mathbf{x}$ is the unique solution of both (P0) and (P1). We will use this proposition as follows: First, we will derive an upper bound on $P_L(\mathbf{D})$ that is explicit in $d$, $b$, and $L$. Second, we will impose that this upper bound falls below $1/2$, and then solve for $L$ to get (8).

Let $\mathbf{v}$ be an arbitrary nonzero vector in $\text{kern}(\mathbf{D})$. Because $\mathbf{D} = [\mathbf{A} \ \mathbf{B}]$, it is convenient to view $\mathbf{v}$ as a concatenation of two vectors $\mathbf{w}$ and $\mathbf{z}$ that satisfy $\mathbf{A}\mathbf{w} = -\mathbf{B}\mathbf{z}$. Let $n$ be the number of elements in $\mathbf{w}$ that belong to the set of $L \geq 1$ largest (in absolute value) entries of $\mathbf{v}$ and let $\mathcal{W}$ be the corresponding index set. Similarly, let $\mathcal{Z}$ be the index set corresponding to the $L - n$ elements of $\mathbf{z}$ that belong to the set of largest (in absolute value) entries of $\mathbf{v}$. Two situations—that we treat separately—may arise: $n > L/2$ and $n \leq L/2$.

*The case $n > L/2$:* We assume, without loss of generality, that[7] $\sum_{i \in \mathcal{W}} |w_i| = 1$. Furthermore, we set $\sum_{i \notin \mathcal{W}} |w_i| = \gamma$, so that $\|\mathbf{w}\|_1 = 1 + \gamma$. Then, we can write $S_L(\mathbf{v})/\|\mathbf{v}\|_1$ as follows:

$$\frac{S_L(\mathbf{v})}{\|\mathbf{v}\|_1} = \frac{\sum_{i \in \mathcal{W}} |w_i| + \sum_{i \in \mathcal{Z}} |z_i|}{\|\mathbf{w}\|_1 + \|\mathbf{z}\|_1} = \frac{1 + \sum_{i \in \mathcal{Z}} |z_i|}{1 + \gamma + \|\mathbf{z}\|_1}$$
$$\leq \frac{1 + (L-n)\|\mathbf{z}\|_\infty}{1 + \gamma + \|\mathbf{z}\|_1}. \tag{20}$$

We next derive an upper bound on $\|\mathbf{z}\|_\infty$ and a lower bound on $\|\mathbf{z}\|_1$ both of which are explicit in $b$ and $d$. The vectors $\mathbf{w}$ and $\mathbf{z}$ satisfy the following equality

$$\mathbf{B}^H \mathbf{A} \mathbf{w} = -\mathbf{B}^H \mathbf{B} \mathbf{z}. \tag{21}$$

To get an upper bound on $\|\mathbf{z}\|_\infty$, we next bound the $\ell_\infty$-norm of both sides of (21). For the LHS we have

$$\|\mathbf{B}^H \mathbf{A} \mathbf{w}\|_\infty \leq \max_{i,j} |[\mathbf{B}^H \mathbf{A}]_{i,j}| \|\mathbf{w}\|_1 \leq d \|\mathbf{w}\|_1. \tag{22}$$

The RHS of (21) can be lower-bounded as follows. Let $\mathbf{G} = \mathbf{B}^H \mathbf{B}$ and note that, by assumption, $[\mathbf{G}]_{i,i} = 1$ and $|[\mathbf{G}]_{i,j}| \leq b$ for $i \neq j$. Then

$$\|\mathbf{G}\mathbf{z}\|_\infty = \max_i \left| z_i + \sum_{j \neq i} [\mathbf{G}]_{i,j} z_j \right| \geq \max_i \left\{ |z_i| - b \sum_{j \neq i} |z_j| \right\}$$
$$= (1+b)\|\mathbf{z}\|_\infty - b\|\mathbf{z}\|_1. \tag{23}$$

Combining (22) and (23), we get

$$(1+b)\|\mathbf{z}\|_\infty - b\|\mathbf{z}\|_1 \leq d\|\mathbf{w}\|_1 \tag{24}$$

or, equivalently,

$$\|\mathbf{z}\|_\infty \leq \frac{d\|\mathbf{w}\|_1 + b\|\mathbf{z}\|_1}{(1+b)} = \frac{d(1+\gamma) + b\|\mathbf{z}\|_1}{(1+b)}. \tag{25}$$

Since the vectors $\mathbf{w}$ and $\mathbf{z}$ also satisfy

$$\mathbf{A}^H \mathbf{A} \mathbf{w} = -\mathbf{A}^H \mathbf{B} \mathbf{z}$$

steps similar to the ones used to arrive at (24) yield

$$(1+a)\|\mathbf{w}\|_\infty - a\|\mathbf{w}\|_1 \leq d\|\mathbf{z}\|_1. \tag{26}$$

Invoking the assumption $a \leq b$, we can lower-bound the LHS of (26) according to $(1+a)\|\mathbf{w}\|_\infty - a\|\mathbf{w}\|_1 \geq (1+b)\|\mathbf{w}\|_\infty - b\|\mathbf{w}\|_1$, which, upon insertion into (26), yields

$$\|\mathbf{z}\|_1 \geq \frac{(1+b)\|\mathbf{w}\|_\infty - b\|\mathbf{w}\|_1}{d} \geq \frac{(1+b)/n - b(1+\gamma)}{d}. \tag{27}$$

In the last step, we used $\|\mathbf{w}\|_\infty \geq 1/n$, as a consequence of the normalization $\sum_{i \in \mathcal{W}} |w_i| = 1$ (recall that $\mathcal{W}$ has cardinality $n$).

Using (25) and (27) in (20), we obtain

$$\frac{S_L(\mathbf{v})}{\|\mathbf{v}\|_1} \leq \frac{1 + (L-n)(d-b)(1+\gamma)/(1+b)}{1 + \gamma + [(1+b)/n - b(1+\gamma)]/d} + \frac{(L-n)b}{1+b}$$
$$= h(n, \gamma, L). \tag{28}$$

---

[7]A scaling of $\mathbf{v}$ does not alter the ratio $S_L(\mathbf{v})/\|\mathbf{v}\|_1$.

This upper bound can be further simplified if we replace $h(n, \gamma, L)$ by its maximum with respect to $n$ and $\gamma$. The function $h(n, \gamma, L)$ is monotonically decreasing in $\gamma$ for $n > L/2$. Hence, we get

$$\max_{L/2 < n \leq L} h(n, \gamma, L) \leq \max_{L/2 < n \leq L} h(n, 0, L) \leq \max_{x \geq 1} h(x, 0, L).$$

For $b = d$, the function $h(x, 0, L)$ does not depend on $x$ and reduces to

$$h(x, 0, L) = Ld/(1+d).$$

Thus, for $b = d$, the inequality $h(x, 0, L) < 1/2$ is satisfied (independently of $x$) if and only if $L < (1 + 1/d)/2$, which coincides with the threshold previously found for general dictionaries in [6], [8].

For $b < d$ and fixed $L$, the function $h(x, 0, L)$ is concave in $x$ for $x \geq 0$. Furthermore, we have that $h(1, 0, L) > \lim_{x \to \infty} h(x, 0, L)$. Denoting the stationary point of $h(x, 0, L)$ by $\tilde{x}_s = \tilde{x}_s(d, b, L)$, the properties of $h(x, 0, L)$ just stated imply that

$$\max_{x \geq 1} h(x, 0, L) = \begin{cases} h(\tilde{x}_s, 0, L), & \text{if } \tilde{x}_s(d, b, L) > 1 \\ h(1, 0, L), & \text{otherwise}. \end{cases}$$

If we now impose that $\max_{x \geq 1} h(x, 0, L) < 1/2$ and solve for $L$, we get the final result (8). The condition $\kappa(d, b) > 1$ in (8) is obtained by solving $h(\tilde{x}_s, 0, L) = 1/2$ and inserting the resulting value of $L$ into the condition $\tilde{x}_s(d, b, L) > 1$.

*The case $n \leq L/2$:* In this case, we repeat the steps detailed above for the case $n > L/2$, swapping the roles of $\mathbf{w}$ and $\mathbf{z}$, and obtain an expression analogous to (28). The resulting threshold coincides with (8).